\documentclass{article}
\usepackage[backend=biber, style=apa, natbib=true, doi=false]{biblatex} 
\usepackage{hyperref}
\usepackage{amsmath}
\usepackage{amssymb}
\usepackage{bm}
\usepackage{graphicx}
\usepackage{xcolor}
\usepackage{fancyvrb}
\usepackage{framed}
\usepackage{makecell}
\usepackage{placeins}
\usepackage[export]{adjustbox}
\NewDocumentCommand{\npautocite}{m}{%
	{\renewcommand{\mkbibparens}[1]{##1}\autocite{#1}}%
}

\numberwithin{equation}{section}

\title{A Comparison of \texttt{R} Packages for Estimating Generalized Linear Mixed Models}
\author{Xiang Li \and Mirko Signorelli}
\date{}

\begin{document}
	\maketitle

	\begin{abstract}
	Generalized linear mixed models (GLMMs) are widely used for analyzing correlated
	data, such as longitudinal and multilevel data. With over 15 \texttt{R} packages
	available on \texttt{CRAN} for fitting GLMMs, practitioners face a difficult choice
	regarding which package yields accurate estimates, converges reliably, and offers
	reasonable computational speed. Existing comparisons are either limited to methods
	within a single package or focus on narrow criteria such as speed alone. To address
	this gap, we systematically compared seven representative \texttt{R} packages---
	\texttt{lme4}, \texttt{GLMMadaptive}, \texttt{glmmTMB}, \texttt{MASS}, \texttt{hglm},
	\texttt{brms}, and \texttt{rstanarm}---that implement different estimation frameworks.
	By using Monte Carlo simulations across 24 scenarios, we evaluated each package in
	terms of convergence ratios, computational time, estimation accuracy, and hypothesis
	testing performance. Our results showed that \texttt{lme4\_AGQ} and \texttt{GLMMadaptive}
	yield the highest accuracy and convergence ratios, although \texttt{GLMMadaptive}
	becomes slower under complex random-effect structures. \texttt{lme4\_LA} and
	\texttt{glmmTMB} are computationally fast but exhibit lower convergence ratios and
	larger bias, especially for variance components. \texttt{MASS} and \texttt{hglm}
	are also fast, but \texttt{MASS} yields liberal univariate tests and \texttt{hglm}
	lacks support for correlated random effects and multivariate testing. Between two
	Bayesian packages, \texttt{rstanarm} converges reliably and produces valid univariate
	tests, whereas \texttt{brms} is extremely slow, limiting its practical utility. Based
	on these findings, we provide practical recommendations for choosing GLMM tool in
	applied research.

	\texttt{R} packages, Monte Carlo simulation
	\end{abstract}

	\section{Introduction}\label{sec:intro}
Generalized linear mixed models (GLMMs) provide a flexible analytical framework for
dependent data and have been widely applied in numerous fields (such as biomedical
research, psychology, and economics) for analyzing longitudinal, multilevel, or spatially
correlated data. GLMMs extend generalized linear models (GLMs), which assume independent
observations, by incorporating random effects that enable the modeling of correlations
between observations. Since the likelihood function of GLMMs has no closed-form
solution, parameter estimation must rely on numerical methods. Common estimation
approaches include penalized quasi-likelihood (PQL; \npautocite{breslow_approximate_1993}),
the Laplace approximation (LA; \npautocite{pinheiro_efficient_2006}), adaptive Gauss-Hermite
quadratures (AGQ; \npautocite{pinheiro_efficient_2006}), hierarchical generalized
linear models (HGLMs; \npautocite{lee_hierarchical_1996}), and Bayesian methods, such
as Markov Chain Monte Carlo (MCMC; \npautocite{zeger_generalized_1991}), variational
inference (VI; \npautocite{kucukelbir_automatic_2015}), and maximum a posteriori (MAP;
\npautocite{bassett_maximum_2019}).\\

To enable the application of these methods to real-world problems, a large number of
\texttt{R} packages have been developed. Based on our review, there are currently over
15 packages available on \texttt{CRAN} for fitting GLMMs. This plethora of \texttt{R}
packages makes a wide range of options available, but it can be a source of confusion
and bewilderment. In practice, statisticians and data scientists are often confronted
with the following questions: which packages yield more accurate parameter estimates?
Which ones are less prone to estimation errors and convergence problems? Which ones
are faster and scale more effectively to large datasets? And, most importantly: which
package(s) should they use when analyzing their data?\\

To date, comparisons of \texttt{R} packages for GLMMs in the literature remain significantly
limited. Some studies have evaluated the performance of different
estimation methods within the same package, without conducting cross‑package comparisons.
For example, \textcite{signorelli_poissontweedie_2021} compared the performance of LA
and AGQ within the \texttt{ptmixed} package. Other studies have compared multiple
\texttt{R} packages, but adopted overly narrow evaluation criteria. For instance,
\textcite{brooks_glmmtmb_2017} compared seven packages in terms of computational speed
and flexibility, but ignored other aspects such as estimation accuracy and hypothesis
testing performance. Consequently, a systematic and comprehensive comparison of
\texttt{R} packages for fitting GLMMs is still lacking.\\

In this study, we address this gap by systematically evaluating selected \texttt{R}
packages for GLMM estimation. After making an inventory of the existing \texttt{R}
packages, we proceed to compare seven \texttt{R} packages that are representative of
different estimation methods using Monte Carlo experiments. We consider several
alternative simulation scenarios, and proceed to systematically evaluate the \texttt{R}
packages in terms of convergence ratios, computational time, estimation accuracy and
hypothesis testing performances.\\

The remainder of this paper is organized as follows. Section~\ref{sec:methods} first
introduces the GLMM and three types of parameter estimation frameworks, along with an
overview of the existing \texttt{R} packages for GLMM estimation. Section~\ref{sec:experiment_design}
describes the data simulation design and evaluation metrics used in the experiments.
Section~\ref{sec:results} presents the results of the simulation experiments. Finally,
Section~\ref{sec:discussion} summarizes the conclusions of the study and discusses its
limitations and future directions.
	\section{Methods}\label{sec:methods}
We consider a dataset comprising longitudinal data from \( n \) subjects, each with
\(m_i\) repeated measurements. Let \( i \in \{1, 2, \dots, n\} \) denote subjects, \(j \in
\{1, 2, \dots, m_i\} \) denote repeated measurements over time for subject \(i\), and
\(Y_{ij}\) denote the response of subject \(i\) at \(j\)th measurement. A GLMM for
the response variable \(Y_{ij}\) can be specified as follows:
\begin{equation*}
	\begin{gathered}
		Y_{ij} | \mathbf{u}_i \sim \mathit{ExpFam}(\mu_{ij}, \phi), \\
		g(\mu_{ij}) = \eta_{ij} = \mathbf{x}_{ij}^T \bm{\beta} + \mathbf{z}_{ij}^T \mathbf{u}_i, \\
		\mathbf{u}_i  \sim \mathcal{N}_r ( \mathbf{0}, \mathbf{\Sigma} ), \\
	\end{gathered}
	\label{eq:glmm_model}
\end{equation*}
where \(Y_{ij} | \mathbf{u}_i\) follows a distribution from the exponential family with mean
\(\mu_{ij}\) and dispersion parameter \(\phi\). The link function \(g(\cdot)\) connects the
expectation \(\mu_{ij}\) to the linear predictor \(\eta_{ij}\). \(\bm{\beta}\) and \(\mathbf{u}_i\)
are vectors of fixed effects of dimension \(k\) and vectors of random effects of
dimension \(r\), respectively. \(\mathbf{x}_{ij}\) and \(\mathbf{z}_{ij}\) are the
corresponding covariates. It is assumed that the random effects \(\mathbf{u}_i\) follow
a multivariate normal distribution with zero mean and covariance matrix \(\mathbf{\Sigma}\).
\subsection{Frameworks for parameter estimation}\label{subsec:frameworks_estimation}
Parameter estimation procedures can be categorized into three types based on the
objective function employed: methods based on marginal likelihood, methods based on
hierarchical likelihood, and methods based on posterior distribution.
\subsubsection{Methods based on marginal likelihood}
The marginal likelihood is the product of the marginal densities of the observed data
\(f(\mathbf{Y}_i | \bm{\beta}, \mathbf{\Sigma})\) over all subjects. Specifically, this marginal density
for a single subject is obtained by integrating the joint density of \(Y_{ij}\) and
\(\mathbf{u}_i\) over the random effects:
\begin{gather}
	\mathcal{L}(\bm{\beta}, \mathbf{\Sigma}) = \prod_{i=1}^n f(\mathbf{Y}_i | \bm{\beta}, \mathbf{\Sigma}) =
	\prod_{i=1}^n \int \left[ \prod_{j=1}^{m_i} f(Y_{ij} | \mathbf{u}_i, \bm{\beta}, \phi)  \right] \cdot f(\mathbf{u}_i | \mathbf{\Sigma}) d \mathbf{u}_i,
	\label{eq:marginal_lkhd}
\end{gather}
where \(f(Y_{ij} | \mathbf{u}_i, \bm{\beta}, \phi)\) is the chosen density from the exponential
family, and \(f(\mathbf{u}_i | \mathbf{\Sigma})\) is the density of the multivariate normal
distribution.\\

The parameters \(\bm{\beta}\) and \(\mathbf{\Sigma}\) can be estimated by maximizing the marginal
likelihood \eqref{eq:marginal_lkhd}. However, for most choices of density function
\(f(Y_{ij} | \mathbf{u}_i, \bm{\beta}, \phi)\) and link function \(g(\cdot)\), the likelihood
involves an intractable integral, making it necessary to resort to numerical integration.
Commonly used numerical integration techniques include PQL \autocite{breslow_approximate_1993},
LA \autocite{pinheiro_efficient_2006}, and AGQ \autocite{pinheiro_efficient_2006}.
This makes it possible to obtain a numeric approximation of the marginal likelihood,
which is then maximized via optimization algorithms. Frequently used optimization algorithms
include, but are not limited to, bound optimization by quadratic approximation (BOBYQA),
non-linear minimization with box constraints (\texttt{nlminb()} in \texttt{R}), and
the Nelder–Mead method.
\subsubsection{Methods based on hierarchical likelihood}
\textcite{lee_hierarchical_1996} extended GLMMs by proposing hierarchical generalized
linear models (HGLMs). HGLMs relax the assumption that random effects must
follow a Gaussian distribution, allowing them to be drawn from a broader family of
distributions. Furthermore, these models allow the variance components of the random
effects to depend on covariates. To estimate the parameters of HGLMs,
\textcite{lee_modelling_2001} introduced a new likelihood function, termed hierarchical
likelihood or h-likelihood, which treats the random effects directly as parameters to be
estimated rather than integrating them out. Differently from the marginal likelihood,
which uses the marginal density of \(\mathbf{Y}_i\), the h-likelihood uses the joint
density of \(\mathbf{Y}_i\) and \(\mathbf{u}_i\):
\begin{gather*}
	\mathcal{L}^h (\bm{\beta}, \mathbf{\Sigma}, \mathbf{u}) = \prod_{i=1}^n f(
	\mathbf{Y}_i, \mathbf{u}_i \mid \bm{\beta}, \mathbf{\Sigma}) = \left[\prod_{i=1}^n \prod_{j=1}^{m_i} f\left(Y_{ij} \mid \mathbf{u}_i, \bm{\beta}, \phi \right)\right]
	\left[\prod_{i=1}^n f(\mathbf{u}_i \mid \mathbf{\Sigma})\right].
\end{gather*}
Rather than maximizing the h-likelihood \(\mathcal{L}^h\) directly, \textcite{lee_modelling_2001}
proposed to employ the extended quasi-likelihood (EQL) method, which involves iteratively
solving a set of extended quasi-likelihood equations to jointly estimate \(\bm{\beta}\),
\(\mathbf{\Sigma}\) and \(\mathbf{u}\).
\subsubsection{Methods based on posterior distribution}
Lastly, Bayesian estimation methods for GLMMs maximize the posterior distribution instead
of the likelihood function. Within this framework, all parameters to be estimated---including
the fixed effects \(\bm{\beta}\), the variance components \(\mathbf{\Sigma}\), and the random effects
\(\mathbf{u}\)---are treated as random variables, and prior distributions are specified
for \(\bm{\beta}\) and \(\mathbf{\Sigma}\) accordingly. Following Bayes’ theorem, the posterior
distribution of the parameters is then derived by combining the likelihood function with
the specified prior distributions \autocite{zeger_generalized_1991}:
\begin{gather*}
	f(\bm{\beta}, \bm{\Sigma}, \mathbf{u} \mid \mathbf{Y}) \propto \left[\prod_{i=1}^n \prod_{j=1}^{m_i} f\left(Y_{ij} \mid \mathbf{u}_i, \bm{\beta}
	\right)
	\right]\left[\prod_{i=1}^n f(\mathbf{u}_i \mid \mathbf{\Sigma})\right] \cdot \pi(\bm{\beta})
	\pi(\bm{\Sigma}),
\end{gather*}
where \(f(\cdot)\) and \(\pi(\cdot)\) represent the probability density functions for the
posterior and prior distributions, respectively. Subsequently, parameter estimates are
obtained using methods such as MCMC \autocite{zeger_generalized_1991}, VI
\autocite{kucukelbir_automatic_2015} or MAP \autocite{bassett_maximum_2019}.
\subsection{\texttt{R} packages}\label{subsec:r_packages}
We conducted an extensive survey of \texttt{CRAN} to identify all currently available
\texttt{R} packages capable of estimating GLMMs. We identified and examined 15 packages.
\begin{itemize}
	\item The packages \texttt{galamm}, \texttt{glmm}, \texttt{GLMMadaptive}, \texttt{glmmEP},
	      \texttt{glmmML}, \texttt{glmmrBase}, \texttt{glmmTMB}, \texttt{lme4}, and \texttt{MASS}
	      maximize the marginal likelihood. See Table~\ref{tbl:R-pack-marginal-lik}
		  for an overview of the numeric integration and optimization methods implemented
		  in these packages
	\item Package \texttt{hglm} maximizes the hierarchical likelihood and applies EQL
		  as the estimation method.
	\item The packages \texttt{bamlss}, \texttt{brms}, \texttt{MCMCglmm}, \texttt{rstanarm},
		  and \texttt{vglmer} are based on the posterior distribution. See Table~\ref{tbl:R-pack-post}
		  for the parameter estimation techniques implemented in these packages.
\end{itemize}

\begin{table}[!htb]
	\centering
	\caption{List of \texttt{R} packages that estimate GLMMs maximizing the marginal
	likelihood. For each package, we list the estimation methods and optimization algorithms
	implemented. L-BFGS-B: limited-memory Broyden-Fletcher-Goldfarb-Shanno bound. BFGS:
	Broyden-Fletcher-Goldfarb-Shanno. CG: conjugate gradient. SANN: simulated annealing.}
	\label{tbl:R-pack-marginal-lik}
	\begin{tabular}{cll}
		\hline
		\textbf{Package} & \textbf{Likelihood approximation} & \textbf{Optimization algorithm(s)} \\
		\hline
		\texttt{galamm} & LA & L-BFGS-B, Nelder-Mead\\
		\texttt{glmm} & Monte Carlo integration & Trust region algorithm\\
		{\bfseries\ttfamily GLMMadaptive} & AGQ &
		\begin{tabular}[l]{@{}l@{}}Nelder-Mead, BFGS, CG, \\L-BFGS-B, SANN, Brent, \\nlminb\end{tabular}\\
		\texttt{glmmEP} & Expectation propagation & BFGS\\
		\texttt{glmmML} & LA, AGQ & BFGS\\
		\texttt{glmmrBase} & MCMC, LA & L-BFGS-B, BOBYQA\\
		{\bfseries\ttfamily glmmTMB} & LA &
		\begin{tabular}[l]{@{}l@{}}Nelder-Mead, BFGS, CG, \\L-BFGS-B, SANN, Brent, \\nlminb\end{tabular}\\
		{\bfseries\ttfamily lme4} & LA, AGQ, PIRLS & BOBYQA, Nelder-Mead\\
		{\bfseries\ttfamily MASS} & PQL & BFGS, L-BFGS-B\\
		\hline
	\end{tabular}
\end{table}

\begin{table}[!htb]
	\centering
	\caption{List of \texttt{R} packages that estimate GLMMs based on the posterior
	distribution. For each package, we list the estimation methods employed.}
	\label{tbl:R-pack-post}
	\begin{tabular}{cl}
		\hline
		\textbf{Package} & \textbf{Estimation method(s)}\\
		\hline
		\texttt{bamlss} & MAP\\
		{\bfseries\ttfamily brms} & MCMC, VI, pathfinder algorithm, MAP\\
		\texttt{MCMCglmm} & MCMC\\
		{\bfseries\ttfamily rstanarm} & MCMC, VI\\
		\texttt{vglmer} & VI\\
		\hline
	\end{tabular}
\end{table}

This extensive array of packages makes it impractical to conduct simulations on all packages and
every method they implement. To keep the comparison manageable, we selected seven of the most
representative \texttt{R} packages for our simulation study. Specifically, for methods
based on marginal likelihood, we selected the two most commonly used packages, i.e.
\texttt{lme4} and \texttt{glmmTMB}, as well as \texttt{MASS} which implements PQL
estimation and \texttt{GLMMadaptive} that provides a comprehensive implementation of
AGQ. For \texttt{lme4}, we included both the LA and AGQ estimation methods. In the
case of hierarchical likelihood, since only the \texttt{hglm} package is available,
it was directly included in the experiment. For methods based on posterior distribution,
we chose \texttt{brms} and \texttt{rstanarm}, both of which were tested using MCMC for
parameter estimation. We employed the default optimization algorithms for each package.\\
	\section{Experiment design}\label{sec:experiment_design}
\subsection{Data simulation}
To evaluate the performance of different packages, we conducted experiments by simulating
data from GLMMs where the response variable is either binary or discrete. In the binary
case, \(Y_{ij} | u_i\) follows a Bernoulli distribution with a logit link function; in the
discrete case, \(Y_{ij} | u_i\) follows a Poisson distribution with a logarithm link function.
We consider three sample sizes \(n = 50, 100, 200\), and assume that individuals can belong
to two groups. Group affiliation is indicated by a dummy variable \(d_i\), where \(d_i
\sim \mathit{Bernoulli}(1,0.4)\). We define a time variable \(t_{ij}\) to represent the
time point of each observation. To enhance the realism of the simulation, we introduced
unbalanced data, where the number of repeated measurements \(m_i\) for each experimental
subject \(i\) varies and follows a discrete uniform distribution \(m_i \sim
\mathit{DU}(1, M)\), with \(M\) being the maximum number of repeated measurements. Two
values of \(M\) are considered in the experiment: \(M = 3\) and \(M = 9\). We define
the time variable \(t_{ij}\) to represent the time point of each observation, where
\(t_{ij}\) is taken from the time set \(\mathbb{T}\). The time set \(\mathbb{T}\) is
specified according to \(M\) in two scenarios: when \(M = 3\), \(\mathbb{T} = \{0,
\frac{1}{3}, \frac{2}{3}, 1\}\); when \(M = 9\), \(\mathbb{T} = \{0, \frac{1}{9},
\frac{2}{9}, \dots, 1\}\). For each subject \(i\), the set of observation time points
\(\{t_{i1}, t_{i2}, \dots, t_{im_i}\}\) is generated by randomly sampling \(m_i\) time
points from \(\mathbb{T}\) without replacement.\\

In the linear predictor, the fixed-effects component is specified as:
\begin{gather}
	\label{eq:fixed_effects}
	\mathbf{x}_{ij}^T \bm{\beta} = \beta_0 + \beta_1 t_{ij} + \beta_2 \mathit{d}_i +
	\beta_3 t_{ij} \times \mathit{d}_i.
\end{gather}

We consider two specifications for the random-effects component: one with only a random intercept
\begin{gather*}
	\mathbf{z}_{ij}^T \mathbf{u}_i = u_{0i}, \\
	\mathbf{\Sigma} = \tau_{0}^2,
\end{gather*}
and one with both a random intercept and a random slope
\begin{gather*}
	\mathbf{z}_{ij}^T \mathbf{u}_i = u_{0i} + u_{1i} t_{ij}, \\
	\mathbf{\Sigma} =
	\begin{bmatrix}
		\tau_{0}^2 & \rho_{01} \tau_{0} \tau_{1} \\
		\rho_{01} \tau_{0} \tau_{1} & \tau_{1}^2
	\end{bmatrix}.
\end{gather*}
Here, \(u_{0i}\) and \(u_{1i}\) represent the random intercept and random slope for
subject \(i\), respectively; \(\tau_{0}\) and \(\tau_{1}\) are their corresponding
standard deviations, and \(\rho_{01}\) is the correlation coefficient between them.\\

\begin{table}[!htb]
	\centering
	\caption{Scenarios considered in the simulation study. This table summarizes
	all possible values of the sample sizes, time sets, response distributions, and
	random-effects structures used in the simulation experiments.}
	\label{tbl:experim_para_set}
	\begin{tabular}{cc}
		\hline
		\textbf{Parameters} & \textbf{Possible values} \\
		\hline
		Sample sizes \(n\) & 50, 100, 200 \\
		Maximum number of repeated measurements \(M\) & 3, 9\\
		Response distributions & Bernoulli, Poisson \\
		Random-effects structures & \begin{tabular}[c]{@{}c@{}}Random intercepts,\\
			random intercepts and slopes\end{tabular} \\
		\hline
	\end{tabular}
\end{table}

In summary, by combining all possible experimental parameters, a total of 24 different
experimental scenarios are obtained (see Table~\ref{tbl:experim_para_set}). For each
scenario, \(S = 1000\) Monte Carlo replications were performed. Furthermore, the true
parameter values are:
\begin{gather*}
	\bm{\beta} = (0.1, 0.3, -0.2, 0.1)^T, \\
	\tau_{0} = 0.8, \tau_{1} = 0.4, \rho_{01} = 0.2.
\end{gather*}
All experiments were conducted on the Academic Leiden Interdisciplinary Cluster Environment
(ALICE, \npautocite{schulz_alice_2022}), running the Red Hat Enterprise Linux 9.5 (Plow)
operating system. The detailed version information for \texttt{R} and the relevant
packages are listed in Table A.1 in Appendix A. The complete code is available at
\url{https://github.com/xanalee/glmmPackCompare}.

\subsection{Performance evaluation}
In Section~\ref{subsec:r_packages}, we introduced the seven \texttt{R} packages that
are compared in our simulation study. To facilitate subsequent discussion, we categorize
five of them—--\texttt{lme4}, \texttt{GLMMadaptive}, \texttt{glmmTMB}, \texttt{MASS} and
\texttt{hglm}--—as frequentist packages, and the remaining two---\texttt{brms} and
\texttt{rstanarm}---as Bayesian packages. Hereafter we introduce the metrics that we
will employ to evaluate both frequentist and Bayesian packages.

\subsubsection{Convergence}\label{subsubsec:conv_prob}
GLMM estimation algorithms often encounter convergence issues. The type and frequency of
those issues varies significantly depending on the estimation method and software implementation
used. Hereafter, we present specific methods that we used to measure the frequency of
convergence problems.\\

For frequentist packages, model convergence refers to whether the algorithm can
successfully find the maximum of the likelihood function. Typically, each package has a
set of built-in convergence checks, which are often quite reliable. However, the built-in
convergence checks in some packages (\texttt{lme4} and \texttt{MASS}) suffer from
reliability issues: even when a check indicates success, signs such as excessively large
parameter estimates may still reveal that the model has not truly converged. Related
discussions can also be found in public forums, for example \textcite{heiko_lme4_2024}.
Therefore, we conducted extra convergence checks on the fitting results from these packages.
The specific procedure is detailed in Appendix A.\\

We define the following four convergence statuses:
\begin{itemize}
	\item \textit{error}: the estimation function returned an error.
	\item \textit{official\_fail}: the estimation function flagged the solution as
	not converged.
	\item \textit{extra\_fail}: the built-in convergence check passed, but our extra
	convergence checks indicated the solution as problematic.
	\item \textit{good}: both the built-in and the extra convergence check passed.
\end{itemize}
\textit{error}, \textit{official\_fail}, \textit{extra\_fail}, and \textit{good} are
mutually exclusive events and collectively comprise all Monte Carlo replications. Based
on the statuses, we define two convergence ratios:
\begin{align}
	\text{Official convergence ratio} &= \frac{\# \mathit{extra\_fail} + \# \mathit{good}}
	{S}, \label{eq:official_conv_ratio} \\
	\text{Real convergence ratio} &= \frac{\# \mathit{good}}{S},
	\label{eq:real_conv_ratio}
\end{align}
where \(\# \mathit{extra\_fail}\) and \(\# \mathit{good}\) denote the number of models that fail extra
diagnostics and the number of models that both pass built-in and extra diagnostics,
respectively.
For packages with reliable built-in convergence checks, namely \texttt{GLMMadaptive},
\texttt{glmmTMB} and \texttt{hglm}, we do not distinguish between the official and real
convergence ratios. \\

For the Bayesian methods considered in our comparison, model convergence refers to
the Markov chain's sample output reaching a stable state. Following the approach
proposed by \textcite{gelman_inference_1992}, we use the \(\hat{R}\) as the criterion
for assessing model convergence:
\begin{gather*}
	\hat{R} = \sqrt{\frac{\frac{N-1}{N} W + \frac{B}{N}}{W}},
\end{gather*}
where \(N\) denote the number of iterations per chain, \(W\) is the within-chain
variance, and \(B\) is the between-chain variance. Therefore, \(\hat{R}\) can be viewed
as the ratio of between-chain to within-chain variance that has been appropriately
adjusted. In the experiments, we adopted four chains, with each chain running 2000
iterations. A model is considered to have adequately converged when \(\hat{R} < 1.1\)
\autocite{gelman_bayesian_1995}.
\subsubsection{Computational time}
Computational time is the most intuitive indicator of a model's computational efficiency.
In the experiments, we use the \texttt{R} function \texttt{system.time()} to record the
fitting time of each Monte Carlo replicate and take the average across all replicates
as the final computation time. Each timing measurement includes only the model fitting
process, excluding subsequent steps such as convergence diagnostics, parameter
extraction, and hypothesis testing.
\subsubsection{Accuracy of estimates}
To evaluate the accuracy of the parameter estimates, we consider two metrics: bias and root
mean squared error (RMSE), formally expressed as follows:
\begin{gather*}
	Bias(\hat{\theta}) = \frac{1}{G} \sum_{s=1}^{G} \hat{\theta}_s - \theta, \\
	RMSE(\hat{\theta}) = \sqrt{\frac{1}{G} \sum_{s=1}^{G} (\hat{\theta}_s - \theta)^2},
\end{gather*}
where \(\theta\) is the true value of the parameter, \(\hat{\theta}_s\) denotes the estimated
value from the \(s\)-th Monte Carlo simulation, and \(G\) is the number of models that
get \texttt{good} in convergence diagnostic.
\subsubsection{Hypothesis testing performance}
To compare the hypothesis testing performance of different methods, we consider both
univariate and multivariate cases. For the model specified in Equation \eqref{eq:fixed_effects}:
\begin{itemize}
\item in the univariate test, the null hypothesis is set as \(H_0: \beta_3 = 0\);
\item in the multivariate test, the null hypothesis is set as \(H_0: \beta_2 = \beta_3 = 0\).
\end{itemize}
For frequentist methods, we employ the Wald test for univariate testing and the
likelihood ratio test (LRT) for multivariate testing. For Bayesian methods, univariate
testing is based on the credible intervals of parameters, while multivariate testing is
evaluated using the leave-one-out information criterion (LOOIC,
\npautocite{vehtari_practical_2017}).\\

We evaluate the testing performance by estimating the empirical type I error and power.
Specifically, if the empirical Type I error rate is significantly higher than the
prespecified level (\(\alpha = 0.05\) in this study), it indicates that the test is too
liberal. Conversely, if it is significantly lower than \(\alpha\), the test is considered
too conservative. A test is deemed reliable when its empirical Type I error rate is
close to \(\alpha\). Meanwhile, a higher empirical power reflects stronger detection capability
of the test provided that the test does control the type I error.
    \section{Results}\label{sec:results}
In this section, we present the experiment results over all 24 scenarios displayed
in Table~\ref{tbl:experim_para_set}. Due to the prohibitive computational cost of
\texttt{brms}, we evaluated it only in the random-intercept-only scenarios. Additionally,
the results for \texttt{lme4} with AGQ are not shown in random-intercept-only scenarios, as
the \texttt{lme4} package does not support models with multidimensional random effects
when using AGQ estimation.

\subsection{Computational time}
\begin{figure}[!htb]
	\centering
	\includegraphics[width=1\textwidth]{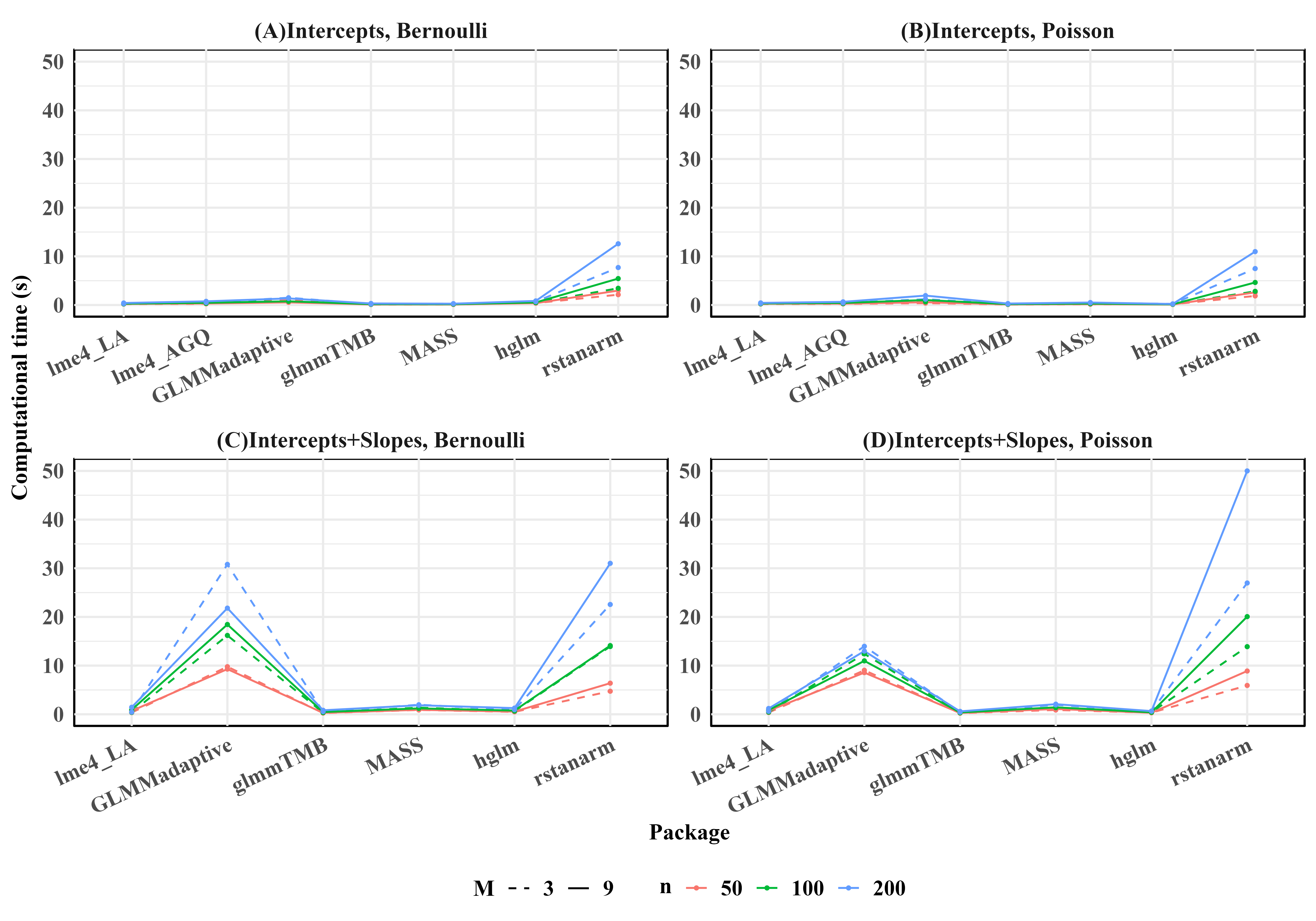}
	\caption{Comparison of computational times across 24 scenarios. Panels (A) and (C)
	correspond to Bernoulli responses, while panels (B) and (D) correspond to Poisson
	responses. Scenarios with only random intercepts are shown in panels (A) and (B),
	and those with both random intercepts and random slopes are shown in panels (C)
	and (D). Different line types distinguish the two maximum numbers of repeated
	measurements: dashed lines represent \(M = 3\), and solid lines represent \(M = 9\).
	Different sample sizes are indicated by red, green, and blue colors. Additionally,
	\texttt{lme4\_LA} and \texttt{lme4\_AGQ} represent parameter estimation using the
	\texttt{lme4} package with LA and AGQ, respectively.}
	\label{fig:computational_time}
\end{figure}

Figure~\ref{fig:computational_time} presents the computational time under different scenarios.
Notably, because the computational time of \texttt{brms} is excessively long, including
it directly in the figure would hinder effective visual comparison among the other packages;
we therefore report the time for \texttt{brms} separately in Appendix B. Comparing the
computation times of the packages, it can be seen that \texttt{lme4}, \texttt{glmmTMB},
\texttt{MASS}, and \texttt{hglm} are very computationally efficient. \texttt{GLMMadaptive}
tends to become slower with two random effects. A plausible explanation is that \texttt{GLMMadaptive}
relies on AGQ for parameter estimation, and its computational complexity increases sharply
with the dimensionality of the random effects. The two Bayesian packages are the slowest,
but \texttt{rstanarm} is considerably faster than \texttt{brms}. \texttt{rstanarm} employs
MCMC as its sampling algorithm, and the iterative nature of this algorithm makes it generally
slower than approximation methods based on numerical integration. This also explains
why \texttt{brms} takes an extremely long time.
\FloatBarrier

\subsection{Convergence ratios}
\begin{figure}[!htb]
	\centering
	\includegraphics[width=1\textwidth]{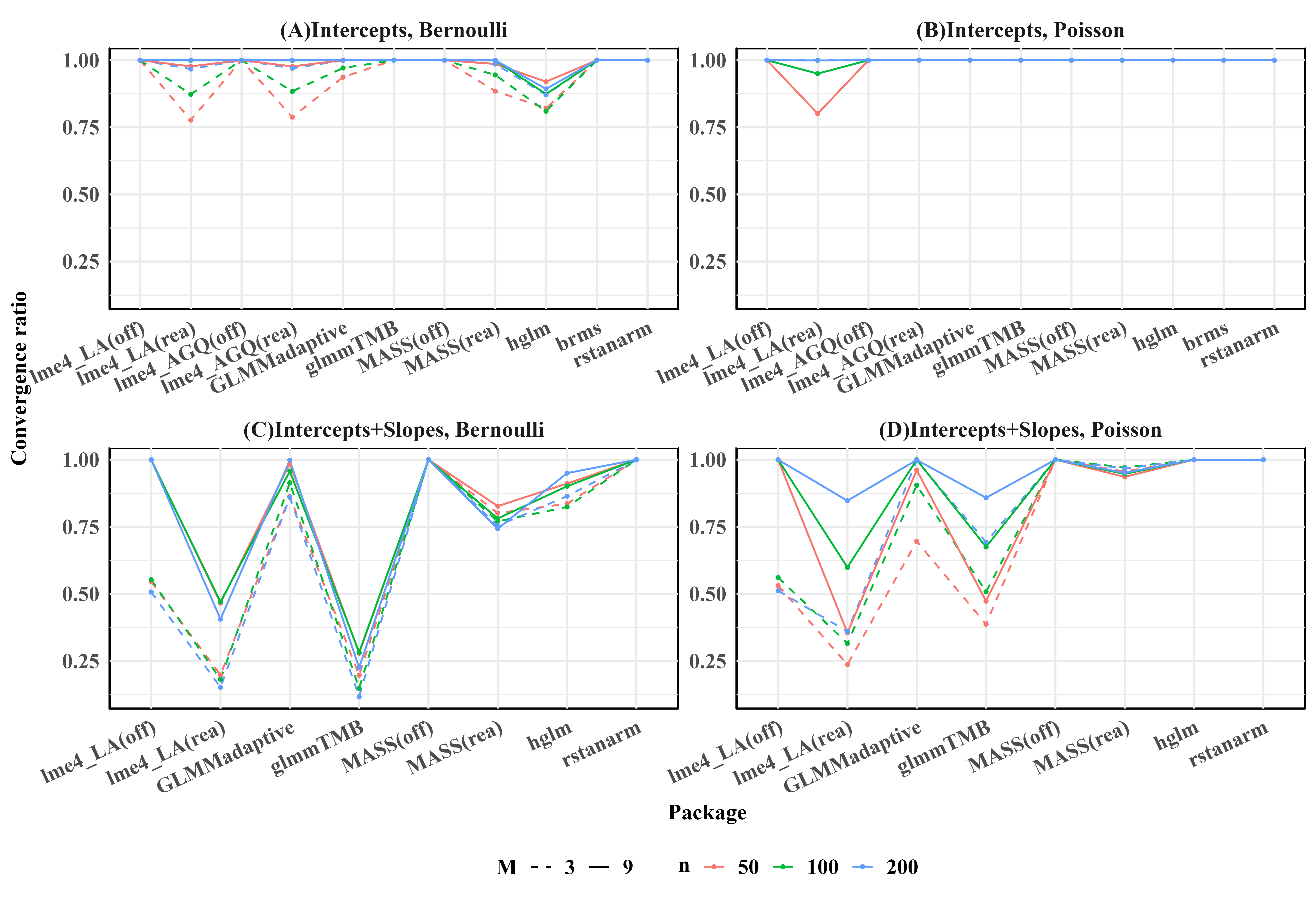}
	\caption{Comparison of convergence ratios across 24 scenarios. The x-axis labels for
	\texttt{lme4}, \texttt{MASS}, and \texttt{hglm} include suffixes: \textit{(off)}
	denotes official convergence ratio as defined by Equation \eqref{eq:official_conv_ratio},
	and \textit{(rea)} denotes real convergence ratio as defined by Equation \eqref{eq:real_conv_ratio}.
	For other figure notes, refer to Figure \ref{fig:computational_time}.}
	\label{fig:convergence_ratios}
\end{figure}

Figure~\ref{fig:convergence_ratios} shows the convergence ratios of each package under
different scenarios. As discussed in Section~\ref{subsubsec:conv_prob}, for \texttt{lme4}
and \texttt{MASS} we defined two types of convergence ratios: official and real convergence
ratios. From Figure~\ref{fig:convergence_ratios}, it can be seen that in most scenarios,
the real convergence ratio is significantly lower than the official convergence ratio,
indicating that the extra convergence checks indeed capture more non-convergence cases.
The subsequent discussion of \texttt{lme4} and \texttt{MASS} will focus exclusively on
the real convergence ratios.\\

Comparing the results of panels (A)+(B) with those of (C)+(D) shows that models containing
both random intercepts and random slopes are more difficult to converge than models with
only random intercepts. This is consistent with the expectation that greater model complexity
leads to more difficult convergence.\\

Comparing the convergence ratios across packages reveals that \texttt{lme4\_LA} and
\texttt{glmmTMB} exhibit relatively low convergence ratios, \texttt{MASS} and \texttt{hglm}
have moderate convergence ratios, whereas \texttt{GLMMadaptive}, \texttt{lme4\_AGQ},
\texttt{rstanarm}, and \texttt{brms} exhibit high convergence ratios. Relating these
results to the estimation methods employed by the packages, one can further conclude
that the convergence ratios of LA (\texttt{lme4\_LA} and \texttt{glmmTMB}) are generally
lower than those of AGQ (\texttt{GLMMadaptive} and \texttt{lme4\_AGQ}) and MCMC (\texttt{rstanarm}
and \texttt{brms}). This aligns with existing findings: \textcite{liu_note_1994}
demonstrated that estimating the likelihood function using multiple quadrature points
makes AGQ converge more easily than LA. The high convergence ratios observed for MCMC
may be attributed to MCMC sampling avoids issues such as boundary estimates that are
commonly encountered in frequentist optimization.
\FloatBarrier

\subsection{Accuracy of estimates}
\begin{table}[!htb]
	\centering
	\caption{Mean absolute bias. For each package and each parameter, the value in the
	table is the average of the absolute biases across multiple scenarios. Colors
	indicate relative performance: \textcolor{green!50!black}{green} means the package has
	lower bias for that parameter, black indicates medium, and \textcolor{red}{red}
	indicates higher. The values in parentheses indicate the true parameter values.
	Because \texttt{lme4\_AGQ} cannot fit models with multidimensional random effects,
	\texttt{hglm} does not support correlations between random effects, and the
	experiments on \texttt{brms} were restricted to the random-intercept-only scenarios,
	the bias of \(\tau_1\) or \(\rho_{01}\) under the above settings is marked by the
	invalid value "-".}
	\label{tbl:mean_abs_bias}
		\begin{tabular}{lccccccc}
		\hline
		\textbf{Package} & $\boldsymbol{\beta_0}$ & $\boldsymbol{\beta_1}$ & $\boldsymbol{\beta_2}$ & $\boldsymbol{\beta_3}$ & $\boldsymbol{\tau_0}$ & $\boldsymbol{\tau_1}$ & $\boldsymbol{\rho_{01}}$ \\
		& (0.1) & (0.3) & (-0.2) & (0.1) & (0.8) & (0.4) & (0.2)\\
		\hline
		\texttt{lme4\_LA} & 0.011 & \textcolor{red}{0.031} & \textcolor{red}{0.024} & \textcolor{red}{0.039} & 0.086 & \textcolor{green!50!black}{0.295} & \textcolor{green!50!black}{0.083} \\
		\texttt{lme4\_AGQ} & \textcolor{green!50!black}{0.007} & \textcolor{green!50!black}{0.011} & 0.015 & 0.020 & \textcolor{green!50!black}{0.036} & - & - \\
		\texttt{GLMMadaptive} & \textcolor{green!50!black}{0.006} & 0.017 & \textcolor{green!50!black}{0.013} & \textcolor{green!50!black}{0.016} & \textcolor{green!50!black}{0.071} & 0.380 & 0.241 \\
		\texttt{glmmTMB} & 0.014 & \textcolor{red}{0.067} & 0.017 & \textcolor{red}{0.039} & \textcolor{red}{0.174} & \textcolor{red}{0.615} & \textcolor{red}{0.280} \\
		\texttt{MASS} & \textcolor{red}{0.038} & 0.020 & \textcolor{green!50!black}{0.013} & 0.017 & 0.099 & \textcolor{red}{0.418} & \textcolor{red}{0.303} \\
		\texttt{hglm} & \textcolor{red}{0.062} & 0.030 & 0.016 & \textcolor{green!50!black}{0.014} & \textcolor{red}{0.289} & \textcolor{green!50!black}{0.175} & - \\
		\texttt{brms} & 0.015 & \textcolor{green!50!black}{0.016} & \textcolor{red}{0.024} & 0.029 & 0.080 & - & - \\
		\texttt{rstanarm} & 0.010 & 0.021 & \textcolor{green!50!black}{0.013} & 0.018 & 0.082 & 0.404 & \textcolor{green!50!black}{0.232} \\
		\hline
		\end{tabular}
\end{table}

\begin{table}[!htb]
	\centering
	\caption{Mean RMSE. For each package and parameter, we report the averages of the
	RMSEs across multiple scenarios. Colors indicate relative performance: \textcolor{green!50!black}{green}
	means the package has lower RMSE for that parameter, black indicates medium, and
	\textcolor{red}{red} indicates higher. The values in parentheses indicate the true
	parameter values. Because \texttt{lme4\_AGQ} cannot fit models with multidimensional
	random effects, \texttt{hglm} does not support correlations between random effects,
	and the experiments on \texttt{brms} were restricted to the random-intercept-only
	scenarios, the bias of \(\tau_1\) or \(\rho_{01}\) under the above settings is marked by
	the invalid value "-".}
	\label{tbl:mean_rmse}
		\begin{tabular}{lccccccc}
		\hline
		\textbf{Package} & $\boldsymbol{\beta_0}$ & $\boldsymbol{\beta_1}$ & $\boldsymbol{\beta_2}$ & $\boldsymbol{\beta_3}$ & $\boldsymbol{\tau_0}$ & $\boldsymbol{\tau_1}$ & $\boldsymbol{\rho_{01}}$ \\
		& (0.1) & (0.3) & (-0.2) & (0.1) & (0.8) & (0.4) & (0.2)\\
		\hline
		\texttt{lme4\_LA} & \textcolor{red}{0.265} & \textcolor{red}{0.366} & \textcolor{red}{0.425} & \textcolor{red}{0.660} & \textcolor{red}{0.510} & \textcolor{red}{1.531} & \textcolor{red}{0.689} \\
		\texttt{lme4\_AGQ} & \textcolor{green!50!black}{0.246} & \textcolor{green!50!black}{0.332} & 0.393 & \textcolor{green!50!black}{0.534} & \textcolor{green!50!black}{0.192} & - & - \\
		\texttt{GLMMadaptive} & 0.254 & 0.360 & 0.404 & 0.563 & 0.272 & 0.609 & 0.608 \\
		\texttt{glmmTMB} & \textcolor{red}{0.340} & \textcolor{red}{0.552} & \textcolor{red}{0.542} & \textcolor{red}{0.847} & \textcolor{red}{0.771} & \textcolor{red}{2.107} & \textcolor{green!50!black}{0.590} \\
		\texttt{MASS} & 0.248 & 0.341 & \textcolor{green!50!black}{0.382} & 0.536 & 0.269 & 0.690 & \textcolor{red}{0.684} \\
		\texttt{hglm} & \textcolor{green!50!black}{0.239} & \textcolor{green!50!black}{0.307} & \textcolor{green!50!black}{0.356} & \textcolor{green!50!black}{0.490} & 0.334 & \textcolor{green!50!black}{0.254} & - \\
		\texttt{brms} & 0.258 & 0.352 & 0.417 & 0.569 & \textcolor{green!50!black}{0.233} & - & - \\
		\texttt{rstanarm} & 0.248 & 0.349 & 0.395 & 0.552 & 0.236 & \textcolor{green!50!black}{0.502} & \textcolor{green!50!black}{0.350} \\
		\hline
		\end{tabular}
\end{table}

Table~\ref{tbl:mean_abs_bias} shows the mean absolute biases of various packages for
different parameters averaged over all scenarios. The detailed results for each scenario are
provided in Figures B.1 - B.7 in Appendix B. By comparing the biases of fixed-effect
parameters and random-effect variance components, it can be seen that the biases of the
variance component parameters are markedly higher than those of the fixed effects. In particular, the
biases for $\tau_1$ and $\rho_{01}$, which are associated with the random slope, are especially
pronounced, with magnitudes that are often comparable to the true parameter values.
Table~\ref{tbl:mean_rmse} presents the mean RMSE across all scenarios for each package and
each parameter. The detailed RMSE for each scenario is shown in Figures B.8 - B.14 in
Appendix B. Comparing the RMSEs with the true parameter values reveals that almost all
estimates are accompanied by considerable variability. The RMSEs for the interaction
coefficient $\beta_3$ is particularly notable, reaching as much as 5 to 8 times its true value.\\

Observing Tables~\ref{tbl:mean_abs_bias} and~\ref{tbl:mean_abs_bias}, we can notice
that estimation accuracy showed significant differences across packages.
\texttt{GLMMadaptive}, \texttt{lme4\_AGQ}, and \texttt{hglm} had lower bias, whereas
\texttt{glmmTMB}, \texttt{lme4\_LA}, and \texttt{MASS} had higher bias; meanwhile, the
estimates from \texttt{hglm} and \texttt{lme4\_AGQ} exhibited smaller fluctuations,
while those from \texttt{lme4\_LA} and \texttt{glmmTMB} showed greater variability.\\

These findings are further corroborated by the detailed results in Figures B.1-B.14.
In addition, these plots reveal that, for most parameters, models with both random
intercepts and random slopes are more prone to high bias and large variability than
models with only random intercepts, and that smaller numbers of observations (small
\(n\) and \(M\)) tend to yield larger bias and variability.
\FloatBarrier

\subsection{Hypothesis testing performance}
\begin{figure}[!htb]
	\centering
	\includegraphics[width=1\textwidth]{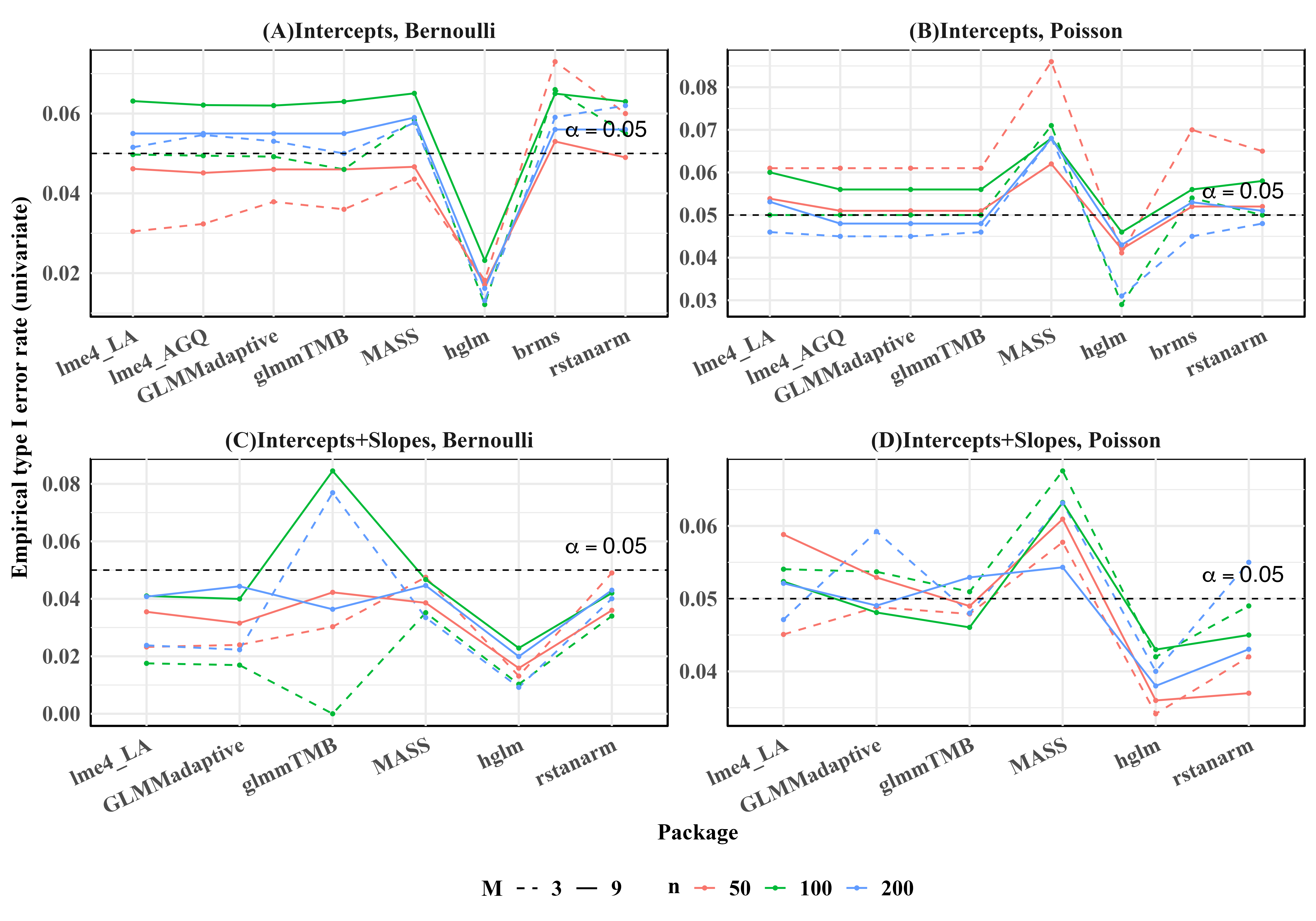}
	\caption{Comparison of empirical Type I error rates for univariate tests across
	24 scenarios. The black dashed line represents the nominal Type I error rate, which
	is 5\% here. For other annotations, refer to Figure \ref{fig:computational_time}.}
	\label{fig:alpha_hat_univariate}
\end{figure}

\begin{figure}[!htb]
	\centering
	\includegraphics[width=1\textwidth]{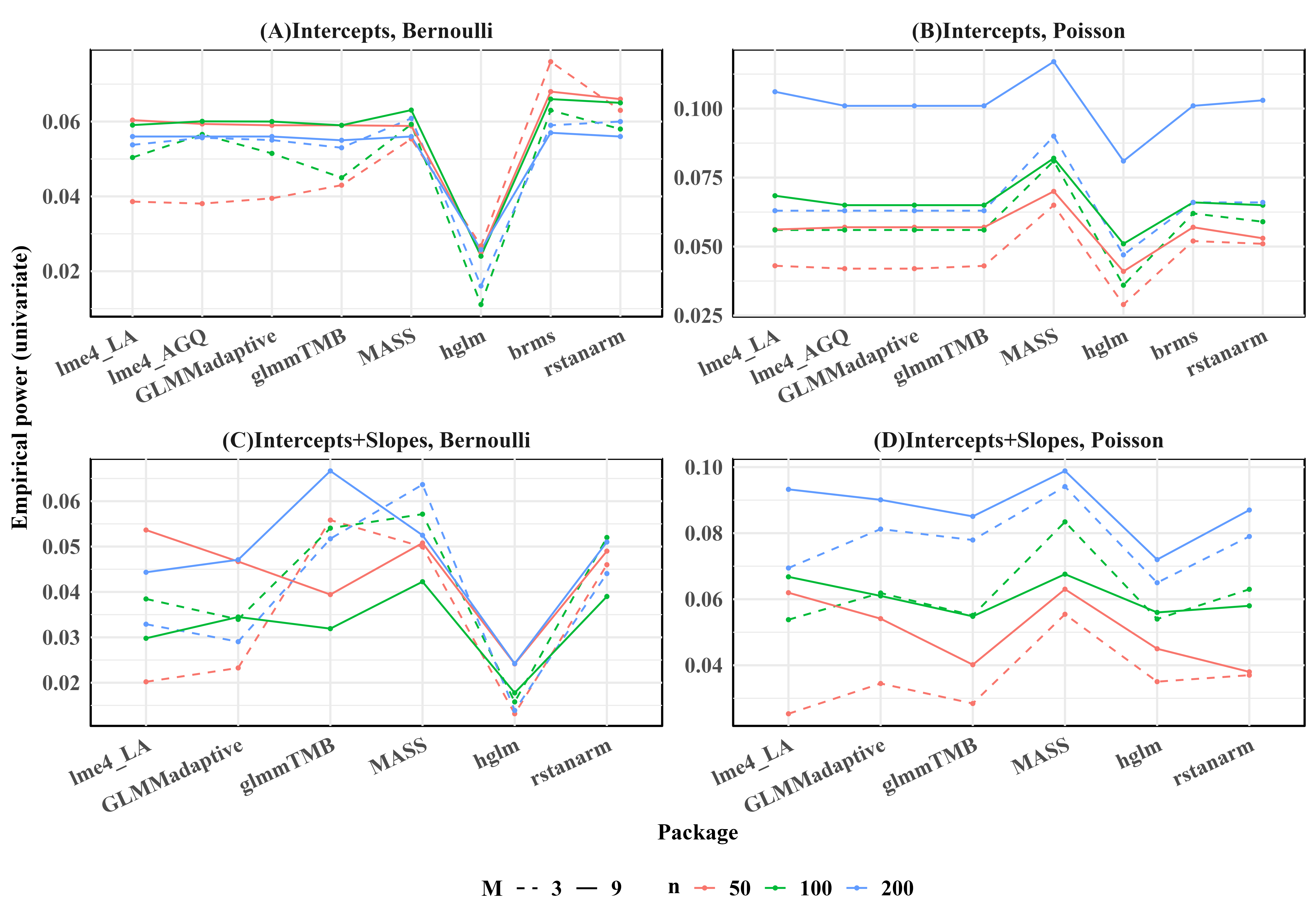}
	\caption{Comparison of empirical power for univariate tests across 24 scenarios.
	For other legends, see Figure \ref{fig:computational_time}.}
	\label{fig:power_hat_univariate}
\end{figure}

\begin{figure}[!htb]
	\centering
	\includegraphics[width=1\textwidth]{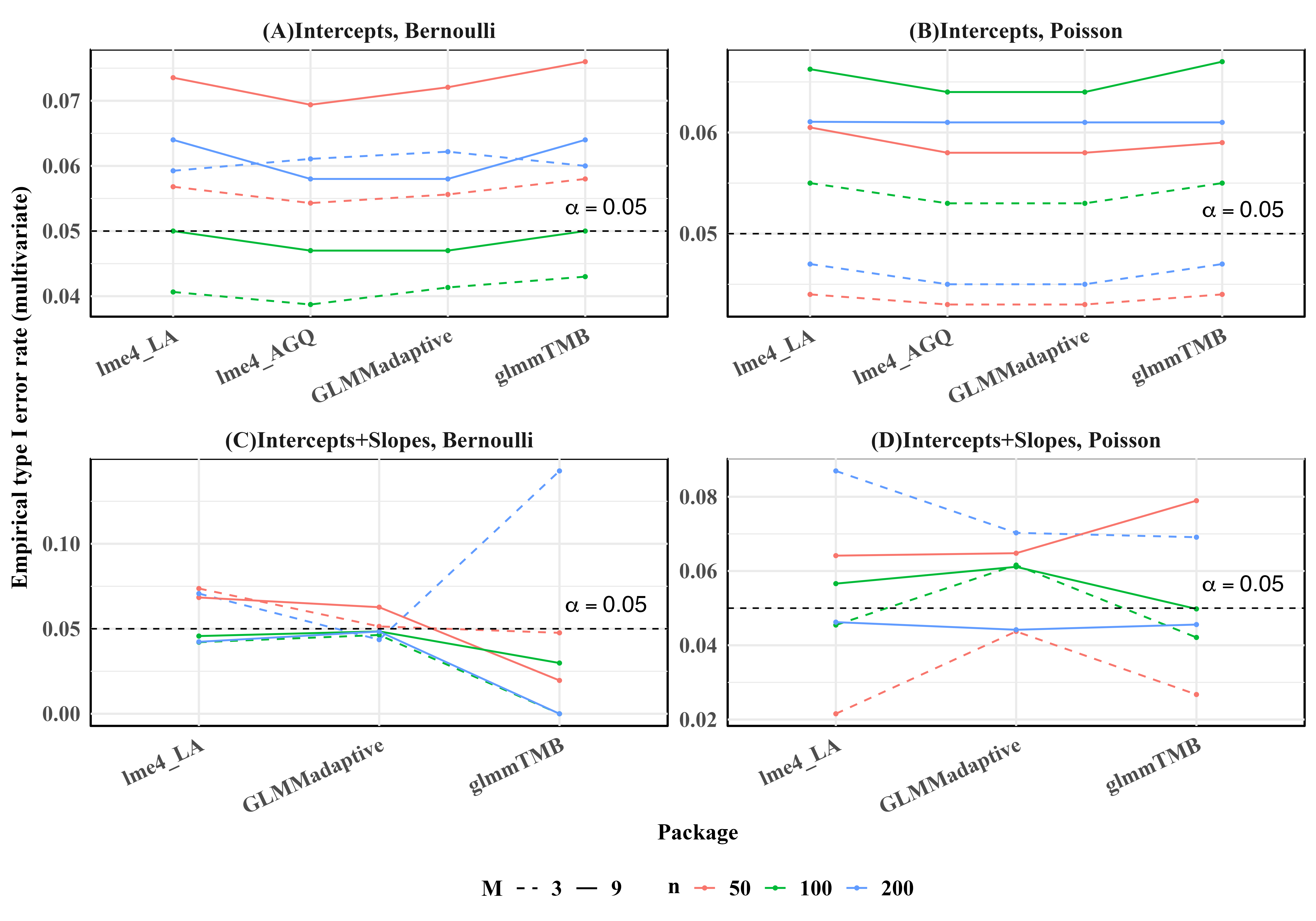}
	\caption{Comparison of empirical Type I error rates for multivariate tests
	across 24 scenarios. \texttt{MASS} and \texttt{hglm} are excluded because they
	lack built-in multivariate tests. \texttt{brms} and \texttt{rstanarm} are
	excluded because they employ a multivariate test based on LOOIC; their performance
	is presented separately in Table \ref{tbl:multi_test_bayes}. The black dashed line
	indicates the nominal Type I error rate, which is 5\% here. For other annotations,
	see Figure \ref{fig:computational_time}.}
	\label{fig:alpha_hat_multivariate}
\end{figure}

\begin{figure}[!htb]
	\centering
	\includegraphics[width=1\textwidth]{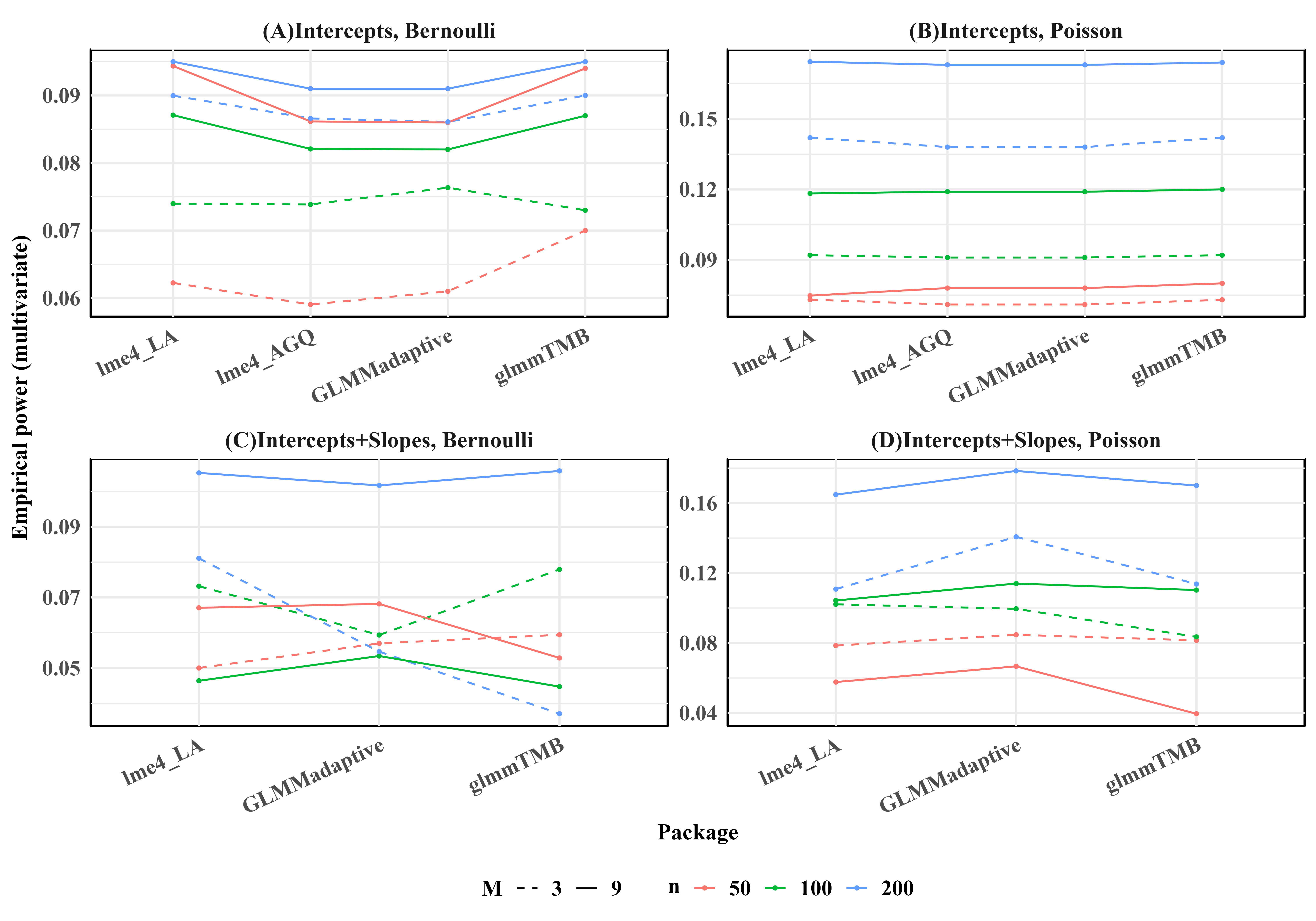}
	\caption{Comparison of empirical power for multivariate tests across 24 scenarios.
	\texttt{MASS} and \texttt{hglm} are excluded because they lack built-in multivariate
	tests. \texttt{brms} and \texttt{rstanarm} are excluded because they employ a
	multivariate test based on LOOIC; their performance is presented separately in
	Table \ref{tbl:multi_test_bayes}. For other legends, see Figure \ref{fig:computational_time}.}
	\label{fig:power_hat_multivariate}
\end{figure}

\begin{table}[!htb]
\centering
\caption{Multivariate test performance of \texttt{brms} and \texttt{rstanarm}.}
\label{tbl:multi_test_bayes}
\begin{tabular}{rcl@{\hspace{0pt}}llc@{\hspace{0pt}}c}
	\hline
	\makecell[c]{$\mathbf{n}$} & \makecell[c]{$\mathbf{M}$} & \textbf{\makecell[c]{Response\\distribution}} & \textbf{\makecell[c]{Random effects}} & \textbf{\makecell[c]{Packages}} & \textbf{\makecell[c]{Type I error\\rate}} & \textbf{\makecell[c]{Power}} \\
	\hline
	50 & 3 & Bernoulli & Intercepts & \texttt{brms} & 0.102 & 0.094 \\
	50 & 3 & Poisson & Intercepts & \texttt{brms} & 0.190 & 0.212 \\
	50 & 9 & Bernoulli & Intercepts & \texttt{brms} & 0.128 & 0.162 \\
	50 & 9 & Poisson & Intercepts & \texttt{brms} & 0.185 & 0.186 \\
	100 & 3 & Bernoulli & Intercepts & \texttt{brms} & 0.096 & 0.138 \\
	100 & 3 & Poisson & Intercepts & \texttt{brms} & 0.203 & 0.236 \\
	100 & 9 & Bernoulli & Intercepts & \texttt{brms} & 0.147 & 0.176 \\
	100 & 9 & Poisson & Intercepts & \texttt{brms} & 0.220 & 0.248 \\
	200 & 3 & Bernoulli & Intercepts & \texttt{brms} & 0.130 & 0.185 \\
	200 & 3 & Poisson & Intercepts & \texttt{brms} & 0.270 & 0.321 \\
	200 & 9 & Bernoulli & Intercepts & \texttt{brms} & 0.152 & 0.208 \\
	200 & 9 & Poisson & Intercepts & \texttt{brms} & 0.269 & 0.349 \\
	\hline
	50 & 3 & Bernoulli & Intercepts & \texttt{rstanarm} & 0.120 & 0.133 \\
	50 & 3 & Bernoulli & Intercepts+slopes & \texttt{rstanarm} & 0.170 & 0.174 \\
	50 & 3 & Poisson & Intercepts & \texttt{rstanarm} & 0.183 & 0.223 \\
	50 & 3 & Poisson & Intercepts+slopes & \texttt{rstanarm} & 0.209 & 0.226 \\
	50 & 9 & Bernoulli & Intercepts & \texttt{rstanarm} & 0.148 & 0.171 \\
	50 & 9 & Bernoulli & Intercepts+slopes & \texttt{rstanarm} & 0.147 & 0.163 \\
	50 & 9 & Poisson & Intercepts & \texttt{rstanarm} & 0.187 & 0.221 \\
	50 & 9 & Poisson & Intercepts+slopes & \texttt{rstanarm} & 0.210 & 0.223 \\
	100 & 3 & Bernoulli & Intercepts & \texttt{rstanarm} & 0.120 & 0.150 \\
	100 & 3 & Bernoulli & Intercepts+slopes & \texttt{rstanarm} & 0.159 & 0.193 \\
	100 & 3 & Poisson & Intercepts & \texttt{rstanarm} & 0.216 & 0.238 \\
	100 & 3 & Poisson & Intercepts+slopes & \texttt{rstanarm} & 0.260 & 0.293 \\
	100 & 9 & Bernoulli & Intercepts & \texttt{rstanarm} & 0.166 & 0.183 \\
	100 & 9 & Bernoulli & Intercepts+slopes & \texttt{rstanarm} & 0.146 & 0.160 \\
	100 & 9 & Poisson & Intercepts & \texttt{rstanarm} & 0.212 & 0.247 \\
	100 & 9 & Poisson & Intercepts+slopes & \texttt{rstanarm} & 0.256 & 0.313 \\
	200 & 3 & Bernoulli & Intercepts & \texttt{rstanarm} & 0.130 & 0.189 \\
	200 & 3 & Bernoulli & Intercepts+slopes & \texttt{rstanarm} & 0.151 & 0.189 \\
	200 & 3 & Poisson & Intercepts & \texttt{rstanarm} & 0.265 & 0.338 \\
	200 & 3 & Poisson & Intercepts+slopes & \texttt{rstanarm} & 0.320 & 0.401 \\
	200 & 9 & Bernoulli & Intercepts & \texttt{rstanarm} & 0.155 & 0.208 \\
	200 & 9 & Bernoulli & Intercepts+slopes & \texttt{rstanarm} & 0.149 & 0.223 \\
	200 & 9 & Poisson & Intercepts & \texttt{rstanarm} & 0.259 & 0.349 \\
	200 & 9 & Poisson & Intercepts+slopes & \texttt{rstanarm} & 0.296 & 0.399 \\
	\hline
\end{tabular}
\end{table}

Figures~\ref{fig:alpha_hat_univariate} and~\ref{fig:power_hat_univariate} present the
empirical type I error rates and empirical power of the univariate tests, respectively.
It can be seen that when the number of observations is small, the empirical type I error
rate tends to deviate from the nominal level; as the number of observations increases,
the empirical type I error rate approaches the nominal value in most cases. Comparing
the performance of different \texttt{R} packages, the univariate tests in \texttt{hglm}
are relatively conservative, whereas those in \texttt{MASS} and \texttt{brms} are
relatively liberal. For \texttt{lme4\_LA}, \texttt{lme4\_AGQ}, \texttt{GLMMadaptive},
\texttt{glmmTMB}, and \texttt{rstanarm}, the empirical type I error rates of the
univariate tests remain around the nominal level, confirming the statistical validity
of the univariate tests in these packages. From the empirical power of the univariate
tests for each package shown in Figure~\ref{fig:power_hat_univariate}, \texttt{MASS}
and \texttt{brms} exhibit higher power, while \texttt{hglm} shows lower power. However,
considering that the tests in \texttt{MASS} and \texttt{brms} are relatively liberal,
their higher power may result from excessive rejection.\\

Figures~\ref{fig:alpha_hat_multivariate} and~\ref{fig:power_hat_multivariate} present
the empirical type I error rates and empirical power of the multivariate tests for the
frequentist packages, respectively. Figure~\ref{fig:alpha_hat_multivariate} reveals that
the empirical type I error rates of \texttt{lme4\_LA}, \texttt{lme4\_AGQ}, \texttt{GLMMadaptive}, and \texttt{glmmTMB} all
remain around the nominal level, confirming the statistical validity of their multivariate
tests. And their empirical powers also show little difference (Figure~\ref{fig:power_hat_multivariate}).\\

Since the multivariate tests in the Bayesian packages \texttt{brms} and \texttt{rstanarm}
are based on LOOIC, which performs model selection by comparing information criteria
rather than hypothesis testing, we did not directly compare the test performance of
frequentist and Bayesian packages. Table~\ref{tbl:multi_test_bayes} separately
presents the multivariate test results for these two packages. Comparing the multivariate
test performance of \texttt{brms} and \texttt{rstanarm}, it indicates that
\texttt{rstanarm} leads to more rejections than \texttt{brms} both under \(H_0\) and
\(H_1\).

\FloatBarrier
    \section{Discussion}\label{sec:discussion}
Significant gaps remain in simulation-based comparisons of \texttt{R} package performance
in GLMM estimation, and systematic comparative analyses are particularly scarce. To
address this issue, our research selects seven representative \texttt{R} packages from
the many that can fit GLMMs and uses simulation experiments to evaluate their parameter
estimation performance across various scenarios. The aim is to provide a more comprehensive
reference for selecting GLMM analytical tools in applied research.\\

\begin{table}[!htb]
	\centering
	\caption{Comparative performance of the packages. \textcolor{red}{Red} indicates favorable
    performance, black indicates medium performance, and \textcolor{green!50!black}{green} indicates
	unfavorable performance.}
	\label{tbl:rela_perf}
		\makebox[\textwidth][c]{
			\begin{tabular}{lccccc}
				\hline
				\textbf{\makecell[c]{Packages}} & \textbf{\makecell[c]{Computational\\time}} & \textbf{\makecell[c]{Convergence\\ratio}} & \textbf{\makecell[c]{Accuracy}} & \textbf{\makecell[c]{Univariate\\testing}} & \textbf{\makecell[c]{Multivariate\\testing}} \\
				\hline
				\texttt{lme4\_LA} & \textcolor{green!50!black}{short} & \textcolor{red}{low} & \textcolor{red}{inferior} & \textcolor{green!50!black}{valid} & \textcolor{green!50!black}{valid} \\
				\texttt{lme4\_AGQ} & \textcolor{green!50!black}{short} & \textcolor{green!50!black}{high} & \textcolor{green!50!black}{superior} & \textcolor{green!50!black}{valid} & \textcolor{green!50!black}{valid} \\
				\texttt{GLMMadaptive} & medium & \textcolor{green!50!black}{high} & \textcolor{green!50!black}{superior} & \textcolor{green!50!black}{valid} & \textcolor{green!50!black}{valid} \\
				\texttt{glmmTMB} & \textcolor{green!50!black}{short} & \textcolor{red}{low} & \textcolor{red}{inferior} & \textcolor{green!50!black}{valid} & \textcolor{green!50!black}{valid} \\
				\texttt{MASS} & \textcolor{green!50!black}{short} & medium & \textcolor{red}{inferior} & \textcolor{red}{liberal} & \textcolor{red}{unsupported} \\
				\texttt{hglm} & \textcolor{green!50!black}{short} & medium & \textcolor{green!50!black}{superior} & \textcolor{red}{conservative} & \textcolor{red}{unsupported} \\
				\texttt{brms} & \textcolor{red}{long} & \textcolor{green!50!black}{high} & medium & \textcolor{red}{liberal} & uncomparable \\
				\texttt{rstanarm} & medium & \textcolor{green!50!black}{high} & medium & \textcolor{green!50!black}{valid} & uncomparable \\
				\hline
			\end{tabular}
		}
\end{table}

Table~\ref{tbl:rela_perf} summarizes the relative performance of each package on the
five evaluation metrics. As can be seen from the table, \texttt{lme4\_AGQ} and
\texttt{GLMMadaptive}, both employing AGQ, show the best overall performance. However,
between these two packages, since \texttt{lme4\_AGQ} does not support estimating
multidimensional random effects, \texttt{GLMMadaptive} becomes the better choice. It
should be noted, however, that the computational efficiency of \texttt{GLMMadaptive}
may decline markedly as the dimension of the random effects increases.\\

Furthermore, \texttt{lme4\_LA} and \texttt{glmmTMB}, both employing the LA estimation
method, perform well on three evaluation metrics. However, because LA uses only a single
quadrature point to approximate the likelihood, both packages suffer from low convergence
rates and poor parameter precision. This drawback has been confirmed by other studies:
\textcite{liu_note_1994}, \textcite{jin_note_2020}, and \textcite{signorelli_poissontweedie_2021}
all found that, compared with AGQ, LA yields lower convergence rates and worse parameter
precision.\\

Among the frequentist packages, \texttt{MASS} (using PQL) and \texttt{hglm} (using HGLM)
show poorer performance. The unsatisfactory performance of \texttt{MASS} is mainly due
to the linearization inherent in the PQL algorithm, resulting in poor estimates. \texttt{hglm}
has functional limitations: it cannot estimate correlated random effects and lacks
multivariate testing capabilities. Therefore, unless spatial data modeling is required,
we recommend using the better-performing packages mentioned earlier.\\

Between the two packages that implement Bayesian methods, \texttt{rstanarm} outperforms
\texttt{brms}. The most prominent issue with \texttt{brms} is its excessively long running
time. Therefore, if Bayesian methods are selected for estimation, \texttt{rstanarm} is
the preferred tool. However, compared with \texttt{GLMMadaptive}, \texttt{rstanarm}
exhibits lower estimation accuracy.\\

In conclusion, our general recommendation to \texttt{R} users is to use \texttt{GLMMadaptive}.
\texttt{rstanarm} is the recommended alternative for users who prefer a Bayesian estimation
approach. Furthermore, \texttt{lme4}'s AGQ implementation can be a valid alternative
for GLMMs with a single random effect estimated on datasets with a very large sample size.\\

Lastly, it is worth mentioning that when designing this study we purposefully took a
selective approach to ensure its feasibility and keep the comparisons manageable. Several
aspects not considered in this study may be interesting and worth investigating. Firstly,
we compared only seven packages and did not evaluate the impact of different optimizers.
Extending the comparison to additional packages and including different optimizers could
be a potential research direction. Secondly, only two response distributions were tested;
future work could incorporate additional distributions, such as the negative binomial,
to examine package performance in fitting more complex models. Finally, this paper did
not evaluate the packages' capabilities in predicting random effects and responses;
subsequent research could assess prediction accuracy, thereby providing a reference for
applied research involving GLMM prediction.\\

	\section*{Acknowledgements}
	\begin{tabular}{@{}p{3cm}p{\dimexpr\linewidth-3cm\relax}@{}}
		\includegraphics[width=3cm, valign=t]{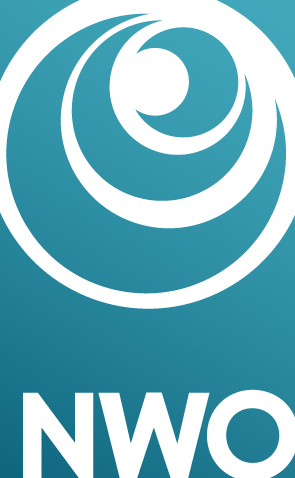} &
		\vspace{0pt}
		This publication is part of the project DynoSurv: flexible statistical models
		for the dynamic prediction of survival outcomes in complex longitudinal studies
		with file number VI.Vidi.233.177 of the research programme Vidi ENW which is
		(partly) financed by the Dutch Research Council (NWO) under the grant
		\url{https://doi.org/10.61686/SSXQW87450}.

		In addition, we acknowledge the ALICE high-performance computing cluster for
		providing computational resources that supported this research.
	\end{tabular}

	\section*{References}
	\printbibliography[heading = none]
\end{document}